\documentclass[superscriptaddress, amsmath, amssymb, aps, sd]{revtex4-2}

\usepackage{graphicx}
\usepackage{dcolumn}
\usepackage{bm}
\usepackage{xcolor}
\usepackage{siunitx}
\usepackage{amsmath}
\usepackage{tabularx}

\begin{document}


\title{\textbf{First experiments with ultrashort, circularly polarized soft X-ray pulses at FLASH2} }%

\author{S. Marotzke}
\affiliation{Deutsches Elektronen-Synchrotron, DESY, Notkestra{\ss}e 85, 22607 Hamburg, Germany}
\affiliation{Institut f\"ur Experimentelle und Angewandte Physik, Christian-Albrechts-Universität zu Kiel, Olshausenstr. 40, 24098 , 24118 Kiel, Germany}

\author{D. Gupta}
\affiliation{Helmholtz-Zentrum Berlin f\"ur Materialien und Energie,  Albert-Einstein-Str. 15, 12489 Berlin, Germany}

\author{R.-P. Wang}
\affiliation{Deutsches Elektronen-Synchrotron, DESY, Notkestra{\ss}e 85, 22607 Hamburg, Germany}

\author{M. Pavelka}
\affiliation{Department of Physics and Astronomy, Uppsala University, Box 256, 751 05 Uppsala, Sweden}

\author{S. Dziarzhytski}
\affiliation{Deutsches Elektronen-Synchrotron, DESY, Notkestra{\ss}e 85, 22607 Hamburg, Germany}

\author{C. von Korff Schmising}
\affiliation{Max Born Institute for Nonlinear Optics and Short Pulse Spectroscopy, Max-Born Stra\ss e 2A, 12489 Berlin, Germany}

\author{S. Jana}
\affiliation{Max Born Institute for Nonlinear Optics and Short Pulse Spectroscopy, Max-Born Stra\ss e 2A, 12489 Berlin, Germany}

\author{N. Thielemann-Kühn}
\affiliation{Freie Universit\"at Berlin, Fachbereich Physik, Arnimallee 14, 14195 Berlin, Germany}

\author{T. Amrhein}
\affiliation{Freie Universit\"at Berlin, Fachbereich Physik, Arnimallee 14, 14195 Berlin, Germany}

\author{M. Weinelt}
\affiliation{Freie Universit\"at Berlin, Fachbereich Physik, Arnimallee 14, 14195 Berlin, Germany}

\author{I. Vaskivskyi}
\affiliation{J. Stefan Institute, Jamova cesta 39, 1000 Ljubljana, Slovenia}

\author{R. Knut}
\affiliation{Department of Physics and Astronomy, Uppsala University, Box 256, 751 05 Uppsala, Sweden}

\author{D. Engel}
\affiliation{Max Born Institute for Nonlinear Optics and Short Pulse Spectroscopy, Max-Born Stra\ss e 2A, 12489 Berlin, Germany}

\author{M. Braune}
\affiliation{Deutsches Elektronen-Synchrotron, DESY, Notkestra{\ss}e 85, 22607 Hamburg, Germany}

\author{M. Ilchen}
\affiliation{Universit\"at Hamburg, Luruper Chaussee 149, 22761 Hamburg, Germany}
\affiliation{Deutsches Elektronen-Synchrotron, DESY, Notkestra{\ss}e 85, 22607 Hamburg, Germany}
\affiliation{Center for Free-Electron Laser Science CFEL, Deutsches Elektronen-Synchrotron DESY, Notkestra{\ss}e 85, 22607 Hamburg, Germany}

\author{S. Savio}
\affiliation{Technische Universit\"at Dortmund, Maria-Goeppert-Mayer-Stra\ss e 2, 44227 Dortmund, Germany}
\affiliation{Universit\"at Hamburg, Luruper Chaussee 149, 22761 Hamburg, Germany}

\author{T. Otto}
\affiliation{Deutsches Elektronen-Synchrotron, DESY, Notkestra{\ss}e 85, 22607 Hamburg, Germany}
\affiliation{Universit\"at Hamburg, Luruper Chaussee 149, 22761 Hamburg, Germany}

\author{K. Tiedtke}
\affiliation{Deutsches Elektronen-Synchrotron, DESY, Notkestra{\ss}e 85, 22607 Hamburg, Germany}

\author{V. Scheppe}
\affiliation{Freie Universit\"at Berlin, Fachbereich Physik, Arnimallee 14, 14195 Berlin, Germany}
\affiliation{Helmholtz-Zentrum Berlin f\"ur Materialien und Energie,  Albert-Einstein-Str. 15, 12489 Berlin, Germany}

\author{J. Rönsch-Schulenberg}
\affiliation{Deutsches Elektronen-Synchrotron, DESY, Notkestra{\ss}e 85, 22607 Hamburg, Germany}

\author{E. Schneidmiller }
\affiliation{Deutsches Elektronen-Synchrotron, DESY, Notkestra{\ss}e 85, 22607 Hamburg, Germany}

\author{C.~Sch\"u\ss ler-Langeheine}
\affiliation{Helmholtz-Zentrum Berlin f\"ur Materialien und Energie,  Albert-Einstein-Str. 15, 12489 Berlin, Germany}

\author{H.~A.~Dürr}
\affiliation{Department of Physics and Astronomy, Uppsala University, Box 256, 751 05 Uppsala, Sweden}

\author{M. Beye}
\affiliation{Stockholm University, SE-106 91 Stockholm, Sweden}
\affiliation{Deutsches Elektronen-Synchrotron, DESY, Notkestra{\ss}e 85, 22607 Hamburg, Germany}

\author{G. Brenner}
\affiliation{Deutsches Elektronen-Synchrotron, DESY, Notkestra{\ss}e 85, 22607 Hamburg, Germany}

\author{N. Pontius}
\email{pontius@helmholtz-berlin.de.}
\affiliation{Helmholtz-Zentrum Berlin f\"ur Materialien und Energie,  Albert-Einstein-Str. 15, 12489 Berlin, Germany}




\date{\today}

\begin{abstract}
Time-resolved absorption spectroscopy as well as magnetic circular dichroism with circularly polarized soft X-rays (XAS and XMCD) are powerful tools to probe electronic and magnetic dynamics in magnetic materials element- and site-selectively. 
Employing these methods, groundbreaking results have been obtained for instance for magnetic alloys, which helped to fundamentally advance the field of ultrafast magnetization dynamics. 
At the free electron laser facility FLASH key capabilities for ultrafast XAS and XMCD experiments have recently improved: 
In an upgrade, an APPLE-III helical afterburner undulator was installed at FLASH2 in September 2023. 
This installation allows for the generation of circularly polarized soft X-ray pulses with a duration of a few tens of femtoseconds covering the $L_{3,2}$-edges of the important $3d$ transition metal elements with pulse energies of several $\mu$J. 
Here, we present first experimental results with such ultrashort X-ray pulses from the FL23 beamline employing XMCD at the $L$-edges of the $3d$ metals, Co, Fe and Ni. 
We obtain significant dichroic difference signals indicating a degree of circular polarization close to 100\%. 
With the pulse-length preserving monochromator at beamline FL23 and an improved pump laser setup, FLASH can offer important and efficient experimental instrumentation for studies on ultrafast spin dynamics in $3d$ transition metals, multilayers, and alloys.
\end{abstract}

\maketitle

\newpage


\section{\label{introduction}Introduction}
 
Soft X-ray based resonant methods have provided one of the biggest contributions to the fundamental understanding of ultrafast magnetization dynamics in multi-element materials within the last two decades. 
The importance hinges on the inherent element specificity of soft X-ray resonances, their sensitivity to the atomic magnetic state, the ability to analyze dissipation of individual spin and orbital moments that determine the materials' magnetization \cite{stammNat.Mater.2007,boeglinNature2010,raduNature2011,wietstrukPhys.Rev.Lett.2011,eschenlohrNat.Mater.2013,henneckePhys.Rev.Lett.2019,goliasPhysRevLett2021,leguyaderAppl.Phys.Lett.2022,janaApplPhysLett2022}, and even the possibility to obtain spatially resolved information \cite{gravesNat.Mater.2013,iacoccaNatCommun2019}. 
These experimental results have significantly advanced the field of magnetization dynamics research.

Soft X-ray methods are element-specific since they probe resonant electronic transitions from localized atomic core levels to valence states. The resonances appear at characteristic, well-separated energies for each element. 
The sensitivity to the atomic magnetic properties, i.e., spin and orbital momenta, is essentially based on the effect of X-ray Magnetic Circular Dichroism (XMCD)\cite{stohrMagnetism:FromFundamentalstoNanoscaleDynamics2006}: the difference in absorption of circularly polarized X-rays of opposite helicity (i.e. photon angular momenta of +1 $\hbar$ or -1 $\hbar$, respectively). 
Selection rules govern the spin-conserving resonant optical transition from a spin-orbit coupled core state into the spin specific density of valence states induced by the magnetic order. This leads to a helicity dependent absorption strength. 
The so-called sum rules can be derived for XMCD spectra, allowing for determination of the absolute atomic orbital and spin magnetic moments for individual elements as well as their dynamic changes \cite{tholePhys.Rev.Lett.1992,carraPhys.Rev.Lett.1993,stohrMagnetism:FromFundamentalstoNanoscaleDynamics2006}.

The crucial requirement for ultrafast dynamics studies with the above mentioned methods is the availability of circularly polarized ultrashort soft X-ray pulses.
First groundbreaking contributions allowed for fundamentally new insights into the ultrafast magnetization dynamics and have already been made by a storage ring based source starting in 2007 \cite{stammNat.Mater.2007,boeglinNature2010,raduNature2011,wietstrukPhys.Rev.Lett.2011,eschenlohrNat.Mater.2013,henneckePhys.Rev.Lett.2019,goliasPhysRevLett2021,leguyaderAppl.Phys.Lett.2022,janaApplPhysLett2022}: 
 Electron bunch slicing was used in combination with an Apple-II type undulator for the generation of circularly polarized soft X-ray radiation at the BESSY II Femtoslicing source \cite{holldackJ.SynchrotronRadiat.2014}. Nowadays, free-electron lasers (FELs) are capable of providing intense X-ray pulses of a few to a few tens of femtoseconds duration or even below. Oftentimes though, the experimental temporal resolution is rather limited by the pulse duration of external pump lasers and the employed synchronization and jitter correction schemes. 

While free-electron laser facilities produce short wavelength radiation since 2005 \cite{ackermannNaturePhoton2007} and soft X-rays since 2009 \cite{Emma_2010}, soft X-rays were initially provided only with linear polarization; ultrashort circularly polarized soft X-ray pulses were produced significantly later at these facilities \cite{lutmanNaturePhoton2016}.

The first FEL based experiments using ultrashort circularly polarized soft X-ray pulses have been reported at the LCLS in 2016 \cite{hartmannRev.Sci.Instrum.2016, higleyReviewofScientificInstruments2016}. 
In 2023, first lasing results with circular polarization at the soft X-ray FEL beamline Athos of SwissFEL  were published \cite{pratNatCommun2023,kittelJSynchrotronRad2024}.
At the FERMI facility ultra short pulses with circular polarization in the XUV, employing an APPLE-II type undulator, were already available since the inception \cite{Allaria_2012}; generation of elliptically polarized femtosecond pulses in a seeded operation mode at the $L_{3,2}$-edges of magnetic transition metals (700 - 800 eV) though has only been demonstrated recently \cite{spezzaniPhys.Rev.B2024}.
At EuXFEL (SA3) an APPLE~X (UE90) undulator for ultrashort X-ray pulse generation with variable polarization has recently entered user operation \cite{yakopovJ.Phys.:Conf.Ser.2022}. 

The increasing number of FEL facilities providing circularly polarized soft X-ray pulses offers new possibilities for ultrafast dynamics studies on magnetic materials. FEL sources have the potential for femtosecond temporal and sub-eV energetic resolutions combined with high X-ray intensities for high fidelity studies. 
In this contribution, we report on the first experiments with ultrashort circularly polarized soft X-ray pulses at FLASH2 in the photon energy range from 700 eV to 860 eV.
We measured XMCD spectra of the $L_{3,2}$-edges of the 
magnetic transition metals Fe, Co, and the $L_3$ edge of Ni using circularly polarized X-ray pulses from the recently implemented afterburner undulator \cite{TischerJ.Phys.:Conf.Ser.2022}.

The availability of circular polarization at the FLASH facility opens up the possibility to extend the available experimental time for soft X-ray methods based ultrafast magnetization dynamics studies. 
The future potential of ultrafast experiments with circularly polarized X-ray pulses on the elemental magnets Fe, Co, and Ni even goes beyond the analytical tools that have been discussed so far. 
Recent studies on the Co $L_3$ edge demonstrate new insight into the ultrafast occupation changes within the \textit{spin-dependent} density of states, governing the magnetic dynamics at different times and timescales \cite{leguyaderAppl.Phys.Lett.2022,pontius2022}.

\section{\label{machine}The FLASH afterburner and radiation diagnostics} 

The measurements reported here were carried out at the XUV \slash soft X-ray free-electron laser FLASH at DESY in Hamburg. FLASH, as a user facility, serves FEL radiation to two different experimental halls \cite{ackermannNaturePhoton2007,Faatz2016}. 
The common FLASH linear accelerator consists of a normal-conducting radiofrequency-gun, super-conducting electron acceleration modules, including a third harmonic cavity, two bunch compressors and a laser heater system. 
It is capable of generating bunch trains with several thousand electron bunches per second in 10 Hz bursts of up to 800 µs length. 
The electron bunches can reach beam energies up to 1.35 GeV and peak currents of up to a few kA. 
A beam distribution yard, based on a kicker and a septum magnet, enables the parallel operation of the two undulator beamlines called FLASH1 and FLASH2. 
The bunch trains can be divided into two variable parts with slightly different properties. 

The FLASH2 undulator beamline contains twelve planar, variable gap undulators, each 2.5 m long.
Downstream from these planar undulators, an APPLE-III type afterburner undulator  was installed at the end of September 2023 \cite{TischerJ.Phys.:Conf.Ser.2022}.
The afterburner undulator allows for generation of FEL radiation with variable polarization at the third harmonic of the FLASH2 SASE FEL radiation, i.e., down to 1.33 nm, correponding to 930~eV photon energy. 
The parameter settings of FLASH2 for the present measurements are given in Table~\ref{parameters}.

The afterburner can be tuned to produce differently polarized X-ray pulses (including circular left and right). 
The main challenge in operating this device is the strong linearly polarized background from the main FLASH2 undulator that contaminates the desired photon polarization state. 
A method for suppression of such a background was proposed in \cite{SchneidmillerPhys.Rev.STAcc2013}: an application of the reverse undulator taper. 
The main effect of this technique is that the radiation is strongly suppressed while the microbunching
at the exit of the main undulator is practically the same as in the reference case of the nontapered undulator.
Then the micro-bunched beam emits powerful radiation with required polarization properties in the afterburner. This method was successfully used to operate the helical afterburner at LCLS \cite{lutmanNaturePhoton2016}.

Since the APPLE-III afterburner at FLASH2 is operated at the third harmonic of the main undulator, the reverse taper method is employed to suppress mostly the weaker linearly-polarized radiation at the third harmonic, which worked very well during the experiment and provided high purity circularly polarized photons. 
The fundamental from the main undulator was also suppressed, although less efficiently than the third harmonic. 
The remaining fudamental radiation was then filtered out in the beamline monochromator, which was used to select the third harmonic. 
This allowed for optimization of the afterburner settings and the experiment itself.  
In addition to applying a reverse taper, a modification of the electron optics at the end of the undulator line was introduced in order to improve the electron beam focusing into the afterburner, thus increasing the radiation intensity and polarization purity.

   \begin{table}[t]
        \centering
        \caption{\label{parameters}
        Overview of the technical parameters of FLASH2 and the pump-laser used in the present experiment.
        \vspace{8pt}
        }
        
        \begin{ruledtabular}
        \begin{tabular}{  r c c  } 
        
           parameter  & FLASH2  & optical laser\\
         \hline
        
        macro train repetition rate     &         10 Hz     &  10 Hz \\
        intra-train repetition rate (spacing)     &         1 MHz (1 $\mu$s)     &  100 kHz  (10 $\mu$s)  \\
        no. of micro pulses            &         400       &  80 (40 used) \\ 
        train duration                  &    400 $\mu$s         &   800 $\mu$s (400 µs used)\\
        wavelength                     & 1.44 - 1.78 nm    & 1030 nm  \\ 
        photon energy                &  700 - 860 eV     &   1.2 eV\\
        electron beam energy            &   1360 MeV        &  n.a. \\
        electron bunch charge           &   340 pC          &  n.a. \\  
        approx. spot size on sample            &  \hspace{1pt} $\sim$~\qtyproduct{12 x 36}{\micro\metre}   & $\sim$~\qtyproduct{175 x 225}{\micro\metre}\\
        
         \hline
        
        FL23 beamline resolution    &  800 meV   (single grating mode) &
        
        \end{tabular}
        \end{ruledtabular}
        \end{table}

The FEL radiation generated with these undulator schemes was characterized by photon-diagnostics measurements during the commissioning of the afterburner undulator prior to the experiment presented here. In this characterization study, we used an instrument comprising a set of circularly arranged photoelectron time-of-flight (TOF) spectrometers to measure the angular distribution of photoelectrons emitted from atomic gas targets in the plane perpendicular to the FEL beam propagation \cite{Braune_2016}. 
In the dipole approximation for a one-photon ionization process, the angular intensity distribution $I$ depending on the polar angle $\theta$ in the detection plane can be described by   

\vspace{-10pt}
\begin{equation}
I(\theta)=\frac{\sigma}{4\pi}\left[1+\frac{\beta}{4}
\left[1+3 \cdot P_\text{lin} \cdot \text{cos}\left(2\left(\theta - \psi\right)\right)\right]\right].\label{eq:AngDistInt}
\end{equation} 
Here, $\sigma$ and $\beta$ are the cross section and the anisotropy parameter, respectively, for the ionization of the particular electronic subshell of the target gas, $P_\text{lin}$ is the degree of linear polarization and $\psi$ the angle of the linear polarization direction with respect to the horizontal axis, i.e. $\psi=0$ represents horizontal linear polarization. While the linear polarization is usually characterized by the Stokes–Poincar\'e parameters $S_{1}$ and $S_{2}$, $P_\text{lin}$ and $\psi$ constitute an alternative representation \cite{Schmidt_1992}: 

\vspace{-10pt}
\begin{equation}
P_\text{lin}=\sqrt{{S_\text{1}}^2 + {S_\text{2}}^2}, \hspace{20pt}                \text{tan}\left(2\psi\right)=\frac{S_\text{2}}{S_\text{1}} \label{eq:altLinPolRep}
\end{equation}

\begin{figure}[t]
\includegraphics[width=\textwidth]{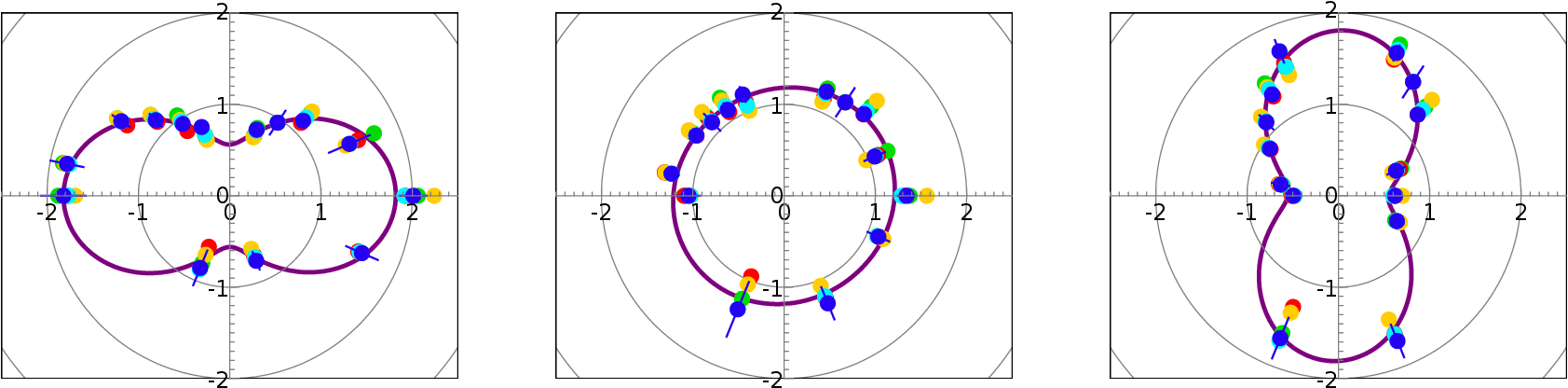}
\caption{ Angular distribution patterns of ionization of Kr $3d$ at $h\nu_{3rd}=313.8$~eV for different polarization modes of the FLASH2 afterburner undulator: linear horizontal (left), circular (mid) and linear vertical (right). The circles represent the measured Kr$\,3d$ photoline intensity for each eTOF and it's corresponding angle in the dipole plane. The colors of the circles refer to different methods to determine the photoelectron signal intensity in the TOF spectrum: photoline profile fit (dark blue), raw intensity integration in a ROI around the photoline with different background definition (light blue, yellow, green) and photoline amplitude values (red). The purple line represents the angular distribution according to Eqn.\ref{eq:AngDistInt} with the weighted mean values for $P_\text{lin}$ and $\psi$.}
\label{fig: AngDistPlots}
\end{figure}

From Eqn.\ref{eq:AngDistInt} it is obvious that with this kind of experimental setup only the linear polarization component of the FEL radiation can be measured directly. In fact, the circular polarization component $P_\text{circ} = S_{3}$ corresponds to an isotropic angular distribution pattern which in principle cannot be distinguished from unpolarized radiation. However, for FELs one can assume almost perfect polarization with a negligible unpolarized component fraction in the order of the Pierce parameter $\rho\thickapprox10^{-3}$ \cite{Geloni_2015}. Hence, the degree of circular polarization can be determined – without being able to differentiate between right and left helicity – by

\vspace{-10pt}
\begin{equation}
P_\text{circ}=\sqrt{ 1 - {P_\text{lin}}^2}.\label{eq:DegreeCP}
\end{equation} 

The capability of this method has been successfully demonstrated previously for different undulator types and polarization setup schemes at FERMI and LCLS \cite{Allaria_2014, lutmanNaturePhoton2016, Ferrari_2015, Hartmann_2016, vonKorffSchmising_2017}.

For the afterburner-undulator characterization at FLASH we used rare gases as ionization targets. Values of ionization cross section $\sigma$ and angular distribution anisotropy parameters $\beta$ for the addressed electronic states of these gases, which are needed for reference, can be found in literature over a wide range of photon energies. Photoelectron angular distribution measurements have been performed for different linear and circular polarization settings of the afterburner undulator at various photon energies of the third harmonic in the range from 74 to 414~eV. Details about the commissioning campaign will be published elsewhere.

As an example, in Figure \ref{fig: AngDistPlots} we present the angular distribution patterns measured at 313.8 eV of the FEL’s third harmonic for three different afterburner undulator settings corresponding to the horizontal linear, vertical linear, and circular polarization. The derived values $P_\text{lin}$ for both linear polarization settings derived from this acquisition data set are $P_\text{lin}^\text{h}=0.96\pm0.02$ and $P_\text{lin}^\text{v}=0.98\pm0.02$, respectively, whereas for circular polarization $P_\text{lin}^\text{c}=0.07\pm0.02$. The uncertainty of 2~\% is the error of the weighted mean of the results using different methods for the photoelectron yield determination, which we applied as a test for a robust analysis procedure suited for a fast diagnostics tool. With a more conservative error estimation the deduced value of the degree of circular polarization of the third harmonic at 313.8~eV is  $P_\text{circ}=99.7 \,^{+0.3}_{-0.4}$~\%.

\section{\label{experiment} Experimental Setup }

In the present measurements, FLASH2 was tuned for the third harmonic radiation to reach the Co and Fe $L_{3,2}$, and the Ni $L_{3}$-edges.
Employing the FLASH2 afterburner undulator (see Section \ref{machine}), circularly polarized photons were generated delivering 400 FEL laser pulses per train.
The train repetition rate was 10 Hz and the intra-train pulse repetition rate was 1 MHz (i.e. 1~$\mu$s single pulse time spacing, total pulse train duration 400~$\mu$s; compare also Table \ref{parameters}).


        \begin{figure}[h]
        \includegraphics[width=14cm]{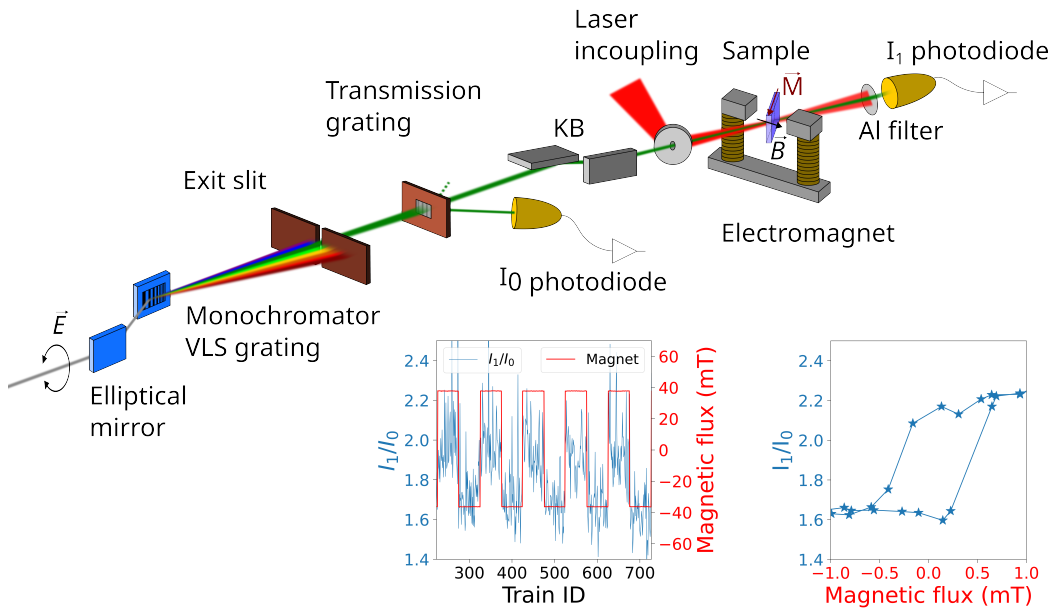}
        \caption{Sketch of the experimental setup. The new FLASH2 helical afterburner undulator enables full polarization control in the photon energy range of the magnetic $3d$ transition metal $L_{3,2}$-edges. For this study, the beamline was operated in a single-diffraction grating mode. A transmission grating behind the monochromator exit slit enables a split-beam normalization scheme for intensity normalization of shot-to-shot SASE fluctuations. The installation of the MUSIX endstation with an in-vacuum electromagnet together with a co-linear laser in-coupling allows time-resolved XMCD measurements. The right hand inset shows a static hysteresis of the FeNi sample at the Fe $L_{3}$-edge; the left hand inset shows representatively the continuously alternating magnetic field together with the recorded normalized transmission signal within a 100 seconds interval of data acquisition.}
        \label{fig: Exp_sketch}
        \end{figure}

The XMCD experiment was carried out at the monochromator beamline FL23 \cite{PolettoJ.Synch.Radiat.2018} of the FLASH2 FEL facility \cite{ackermannNaturePhoton2007,Faatz2016} employing the MUSIX (multidimensional spectroscopy and inelastic x-ray scattering) end-station \cite{BeyeJ.Phys.Condens.Matter_2018}, see Figure \ref{fig: Exp_sketch}. 
Two 3~mm iris apertures were used upstream of FL23 to constrain the beam path. 
The beamline was operated in first-order single-diffraction grating mode to select the third harmonic radiation, using the high-energy grating with a groove density of 600~lines/mm and an energy dispersion of 0.017~nm/mm in the energy range 700~eV to 860~eV. 
An exit slit of 100~$\mu$m width led to an energy resolution of approx. 800~meV. 
In order to maximize the beamline transmission the coatings of all beamline optics were either set to Ni (Fe and Co $L_{2,3}$-edge) or Pt (Ni $L_3$-edge), respectively. 
The first diffraction order of a transmission grating placed after the middle slit of the first FL23 monochromator was detected by a biased photo-diode to act as monitor of the incidence X-ray intensity $I_0$ while the zeroth order was guided further towards the sample \cite{Brenner_2019, Engel_2021}. 
The transmission grating consists of a 100~nm Si$_3$N$_4$ membrane on which grating lines made from a HSQ (hydrogen silsesquioxane) resist with a height of \textgreater~200~nm, a duty cycle of 0.5, and a grating period of 320~nm are fabricated. 
For the zeroth order, a focusing Kirkpatrick-Baez active optics system (KAOS2) was used to focus the beam onto the FeNi or CoPt sample to a transversal spot size of \qtyproduct{12 x 36}{\micro\metre}. 
The intensity $I_1$ of the transmitted beam was measured with a second biased photo-diode. 
The signals of both X-ray photo-diodes were measured for every individual X-ray pulse enabling accurate normalization on a shot-to-shot basis. 
To avoid saturation of the photo-diodes during the measurements at the Fe $L_{2,3}$-edge, the gas attenuator (Krypton) was used at $1.5~\times~10^{-2}$~mbar.

To excite the sample in the time-resolved measurements, a fixed time-structure pulse train of 80 optical laser pulses (1030~nm) at a frequency of 100~kHz (i.e., 10~$\mu$s pulse spacing; pulse train duration 800~$\mu$s) at a train repetition rate of 10 Hz was provided. 
Since the optical laser pulse train showed an almost linear pulse energy build-up for the first 20 pulses before reaching a constant pulse energy level, the delay to the X-ray pulse trains was set such that the first FEL pulse in a train overlapped with laser pulse 21.
Due to the different repetition rates of FEL (1~MHz) and optical laser pulses (100~kHz), every tenth FEL pulse was temporally overlapped with an optical pump laser pulse, such that a total of 40 FEL pulses of a pulse train could be used for the time-resolved measurements.   
The pulse energy of the optical laser pulses were varied by combining two sets of neutral density filters in the beam path. 
The spot size of the optical laser on the sample was separately determined by knife-edge scans (see Table \ref{parameters}) and substantially larger than that of the FEL beam. Spatial overlap of laser and FEL spots was verified by a Ce-doped YAG based fluorescence crystal (imaged with a long-working distance microscope and a CCD camera) at the sample position.  
 The BAM (beam arrival monitor, measuring the electron beam timing relative to a master clock) and the LAM (laser arrival monitor, referencing the laser pulses to the same master clock) were used to correct the data for jitter and drifts between FEL and pump laser \cite{atia-tul-noorOpt.Express2024,lautenschlager:ipac2021-tupab302, schulz:procibic2024-that1}.

The X-ray absorption spectra of the Fe, Co $L_{3,2}$ and Ni $L_{3}$ edges ($L_{2}$ corresponding to the $2p_{{1}/{2}} \rightarrow 3d$, and $L_{3}$ to the $2p_{{3}/{2}} \rightarrow 3d$ transitions, respectively) were obtained by simultaneously moving the  gap of the planar undulators and the helical afterburner together with the monochromator pitch angle.
For the XMCD measurements, the X-ray beam transmitted through the samples at 35-45 degrees with respect to the sample normal, and their magnetization was flipped horizontally in the sample plane by an external magnetic field (see Figure \ref{fig: Exp_sketch}). 
The latter was generated by an in-vacuum electromagnet. 
Magnetic saturation of the samples was ensured by measuring the hysteresis. 
During the measurements, the magnetic field direction was inverted every 10~seconds. 
The circular polarization of the FEL beam was kept fixed for all measurements.

The samples were grown by magnetron sputtering at the Max Born Institute Berlin. The substrates were Si frames (Silson Ltd.) with 15 Silicon-Nitride windows of 0.5mm x 0.5mm lateral size. The layer structure of the CoPt sample was Al(100nm)\hspace{0pt}/\hspace{0pt}SiN(200nm)\hspace{0pt}/\hspace{0pt}Ta(2nm)\hspace{0pt} /Co$_{50}$Pt$_{50}$(25nm)\hspace{0pt}/\hspace{0pt}MgO(2nm), and that of the FeNi sample Al(100nm)\hspace{0pt}/\hspace{0pt}SiN(200nm)\hspace{0pt}/\hspace{0pt}Ta(2nm)\hspace{0pt}/Fe$_{40}$Ni$_{60}$(25nm)\hspace{0pt}/MgO(2nm). The Al layer served as a heat sink, the MgO layer as protection against oxidation.

For analyzing the degree of X-ray polarization at FLASH2, reference XMCD measurements were done after the FLASH2 studies on the identical samples at the PM3 beamline of the BESSY II storage ring \cite{kachelJSynchrotronRad2015}.

\section{\label{results} experimental results  } 

        \begin{figure}[t]
        \includegraphics[width=14cm]{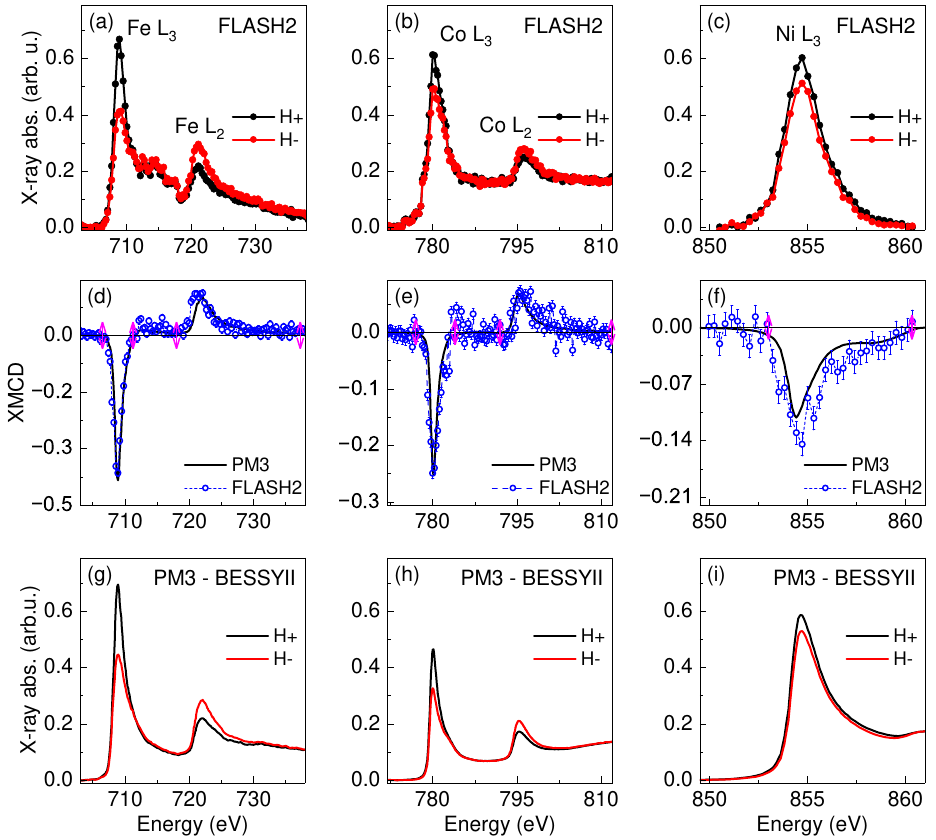}
        \caption{ Static XAS and XMCD spectra at the Fe, Co and Ni $L$-edge:
        (a-c) show XAS spectra recorded at the F23 beamline of FLASH2 with circular polarized X-ray pulses. 
        For Ni, only the $L_{3}$ absorption resonance is shown up to 860 eV, which 
        was the upper photon energy limit of the FLASH2 accelerator during the present measurement campaign. 
        (g-i) XAS spectra of the same samples taken for reference at the BESSY~II storage ring. 
        (d-f)  the XMCD spectra derived from the corresponding XAS spectra.
        The pink arrows indicate the integration limits for the comparison of the absolute value of XMCD in the FL23 and PM3 measurements (see text).
        }
        \label{fig:XMCD_static}
        \end{figure}

Symbols in Figures \ref{fig:XMCD_static}(a-f) shows static XAS and XMCD spectra of the $L_{3,2}$-edge photon energy region of Fe, Co, and Ni measured at FLASH2. XAS spectra for opposite magnetic field orientation (black and red) are presented in plots (a-c). The spectra were calculated from the acquired raw data as: 
\begin{equation}\label{eq1}
T(E) = \frac{I_1}{I_0} 
\end{equation}
\begin{equation}\label{eq2}
XAS(E) = \text{log} (\frac{T(E)} {T(\text{`}E<L_3\text{'})} ) \cdot \text{cos}(\vartheta)
\end{equation}
\begin{equation}\label{eq3}
XMCD(E) = ( XAS^{+M}(E) - XAS^{-M}(E) ) / \text{sin}(\vartheta)
\end{equation}
Here, $T(E)$ is the sample transmitted intensity signal (diode signal $I_1$) normalized to $I_0$ (monitor diode signal) on a shot-to-shot basis.
For proper normalization, events with almost vanishing $I_0$ values were discarded. 
The such obtained X-ray transmission spectra, $T(E)$, were normalized to the signal level in the photon energy range (few eV) immediately below the $L_3$ resonance, denoted as `$E<L_3$'. 
The logarithm of the normalized transmission spectra then scales with the X-ray absorption coefficient of the $L_{3,2}$-edge optical transitions \cite{stohrMagnetism:FromFundamentalstoNanoscaleDynamics2006}.  
For the different resonances we tuned the angles of incidence, $ \vartheta$, to optimize both, the transmission signal by changing the effective sample thickness and hence the value of absorption, and the XMCD amplitude determined by the projection angle of the in-plane sample magnetization onto the X-ray k-vector. In order to compensate for different absorption of various angles of incidence, the XAS spectra where multiplied by the factor $\text{cos} (\vartheta) $. 
These XAS spectra are shown in Figures \ref{fig:XMCD_static} (a-c) for Fe, Co, and Ni, respectively.
Note that for Ni only the $L_3$ absorption edge could be measured, as the $L_2$ absorption edge energy was beyond the limits of the FLASH2 accelerator setting in the present measurement campaign. The net acquisition time for one pair of XAS spectra (H$^+$ and H$^-$)  was between three and four hours.


The XMCD spectra are calculated by subtracting both spectra, $XAS^{+M}(E)$ and $XAS^{-M}(E)$, recorded for opposite magnetic field directions. To correct for the varying projections of the sample magnetization onto the X-ray direction of individual measurements, the XMCD spectra are divided by the factor $\sin(\vartheta)$ (see Eqn.~\ref{eq3}) \cite{stohrMagnetism:FromFundamentalstoNanoscaleDynamics2006} . Hence all XMCD spectra are scaled to the unified configuration with the magnetic field vector parallel to the X-ray direction. The resulting XMCD spectra for Fe, Co, and Ni are shown in Figures \ref{fig:XMCD_static} (d-f) (open blue circles), respectively.  All  elements show significant XMCD. 


We compare these spectra with reference spectra  taken from the identical samples at the PM3 beamline of the BESSY II storage ring (see Figures \ref{fig:XMCD_static} (g-i)) \cite{kachelJSynchrotronRad2015}.  
The latter are derived from the raw data in the same way as those recorded at FLASH2, where here $I_1$ is the photo-current of a photo-diode detecting the sample transmitted intensity, and $I_0$ the  corresponding signal measured without any sample (incident intensity).
For proper comparison, we have to consider that the XMCD spectra for FeNi have been recorded at different temperatures (PM3:~295~K; FL23:~80~K). 
The lower temperature for the  FL23 measurements results in a slightly larger saturation magnetization and hence an increased XMCD. 
Referring to temperature-dependent studies on similar samples from literature \cite{tangJ.Appl.Phys.1996}, we find an $\sim$~8$\%$ larger magnetization for FeNi at 80~K, which we assume to directly translate into the XMCD amplitude. 
This factor is considered in the comparison of the FLASH2 spectra and their PM3 reference spectra in Figures \ref{fig:XMCD_static} (g-i).

        \begin{table}[t]
        \caption{\label{XMCDcomparison}
        The table comprises the absolute values of the energy integration over the XMCD $L_{2,3}$-resonances of Fe, Co, and Ni for the spectra measured at PM3 (row XMCD$_\text{PM3}$) and FL23 (row XMCD$_\text{FL23}$), respectively, given in arbitrary units (compare Figures \ref{fig:XMCD_static} (d-f) and see text for details). 
        The experimental errors given for the XMCD$_\text{FL23}$ values are purely statistical 
        calculated by error progression from the mean error of the XMCD spectra data points.
        The corresponding statistical experimental errors for XMCD$_\text{PM3}$ are negligible and therefore not shown. 
        For the PM3 measurements, the experimentally determined degree of circular polarization, $S_3$, for the respective absorption edges is given underneath the integration numbers  \cite{kachelJSynchrotronRad2015,CitekeyMisc1}.  
        $S_3$ for FL23 is estimated by scaling the PM3 value by the experimentally determined ratio of the integrated XMCD values of the individual resonances. The experimental errors of the $S_3$ values are in the order of 10\% for Fe~$L_3$, $L_2$ and Co $L_3$, and in the order of 25\% for Co~$L_2$ and Ni~$L_3$. 
        \vspace{8pt}
        }
        
        \begin{ruledtabular}
        \begin{tabular}{lccccc}
        
                            &       Fe $L_3$        &       Fe $L_2$        &      Co $L_3$     &       Co $L_2$        &       Ni $L_3$    \\
         \hline
         \hline
         
        XMCD$_\text{PM3}$   &       0.66            &        0.57           &        0.42       &        0.28           &        0.26       \\
        $S_3$ (\%)          &      92.5             &        92.5           &        92.7       &        92.8            &        93.0        \\
        \hline
        XMCD$_\text{FL23}$  &  0.69 $\pm$ 0.050     &    0.67 $\pm$ 0.09   & 0.54 $\pm$ 0.04  &   0.26 $\pm$ 0.08    & 0.39 $\pm$ 0.07   \\
        $S_3$ (\%)          &       97              &         100           &         100       &    86                 &   100             \\
         \hline
        \end{tabular}
        \end{ruledtabular}
        \end{table}

The known degree of circular polarization for the PM3 measurements \cite{kachelJSynchrotronRad2015, CitekeyMisc1} and the unified representation of all XMCD spectra allows us to estimate the degree of circular polarization of FLASH2 from the measured XMCD spectra. 
To correct for the different energy resolution of the two monochromators (F23: 800~meV; PM3: 300~meV), we integrate the XMCD spectra over the individual $L_{2,3}$-resonances. 
The absolute values of this integration are shown in Table \ref{XMCDcomparison} for each resonance (row XMCD$_\text{PM3}$ and XMCD$_\text{FL23}$, respectively).

For the PM3 reference measurements, the degree of the X-ray circular polarization ($S_3$, Stokes–Poincar\'e parameter) has been experimentally determined, or extrapolated from an experimentally determined value, respectively (see Table \ref{XMCDcomparison}) \cite{kachelJSynchrotronRad2015, CitekeyMisc1}. 
Since the samples are identical, and the spectra have been normalized to a unified measuring condition, the XMCD values determined for FL23 allow an assessment of $S_3$ for the FLASH2 measurements by scaling the PM3 $S_3$ value by the ratio of the XMCD. 
The such calculated $S_3$ values of the FLASH2 circular polarization for the individual $L$-resonance energies are given in the bottom row of Table \ref{XMCDcomparison}. Note, that $S_3$ values greater than 100\% are capped at 100\%, as greater values do not make any physical sense. 
For $S_3$, we effectively expect values of almost 100\%. The apparently small value for Co $L_2$ of 86\% is probably due to the relatively large experimental error (see caption of Table \ref{XMCDcomparison}). 
The differing values for Co $L_3$ and $L_2$, however, could as well stem from a non-synchronous gap change of the Apple-III afterburner and the main FLASH2 undulator during the energy scans. 
This is a matter of undulator parameters calibration and will be improved in the future.

To demonstrate the feasibility of ultrafast dynamics magnetic studies with circular polarization, we present the ultrafast element specific XMCD transient of Fe in the FeNi sample after laser excitation. 
Due to the 20 laser pulses in each pulse train, that arrived before the first pulse of the FEL pulse train (intensity build-up, see Section \ref{experiment}), a considerable nearly constant heating of the samples was observed leading to a significant reduction of the sample base magnetization and XMCD.
Even liquid nitrogen cooling of the samples could not remove this effect.
Yet, a compromise had to be found for the pump-laser fluence, minimizing the heating while dynamically inducing a significant demagnetization.

        \begin{figure}[h]
        \includegraphics[width=7cm]{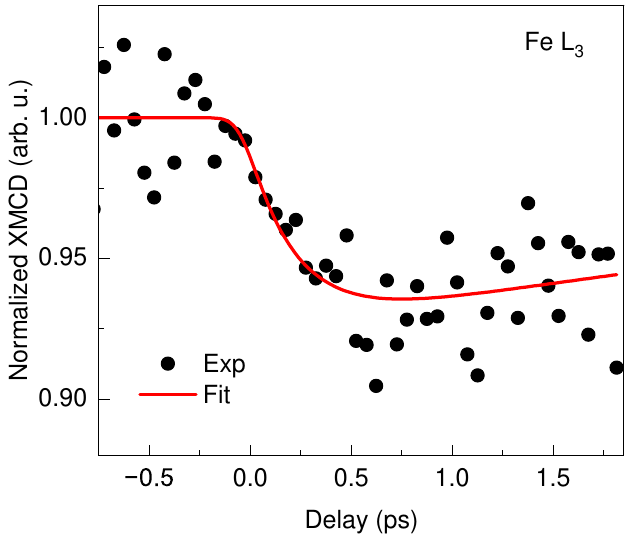}%
        \caption{Normalized transient demagnetization dynamics of Fe in FeNi measured at the Fe $L_{3}$ absorption resonance (black balls). The solid red line represents the exponential fit with Gaussian convolution.}
        \label{fig:dynamic_XMCD_Fe}
        \end{figure}

The effective temporal resolution of this measurement results from various contributions (see Table \ref{timingparameters}): the laser and the X-ray pulse widths and the jitter/drift in the FEL-laser synchronization. 
The latter contributions can partially be corrected for by measurements of BAM and LAM (see Section \ref{experiment}). 
Additional temporal broadening can arise from the beamline monochromator which was used in the single grating mode. 
Here, the X-ray pulse elongation depends on how many grating grooves are illuminated by the X-ray beam and can vary between 100 and 200~fs depending on the beamline alignment. 
Summing up all the contributions, the temporal resolution for our experiment was expected to be in a range between 170 and 270~fs.

        \begin{table}[t]
        \centering
        \caption{\label{timingparameters}
        Overview of the parameters of FLASH2 and the pump laser determining the effective temporal resolution during the present beamtime. The parameters enter the temporal resolution in different manners (see footnotes). The BAM allows to correct the train-to-train jitter from the data. The shot-to-shot jitter/drift of the single X-ray pulses within the train (30~fs) was not corrected in the present measurement and fully enters the effective temporal resolution. BAM correction leaves a remaining broadening of the train-to-train jitter less than 10~fs. Without the elongation by the monochromator, an effective temporal resolution of 90~fs would result.    
        \vspace{8pt}
        }
        \begin{ruledtabular}
        \begin{tabular}[b]{l c} 
            parameter  &  value \\
             \hline
         
             laser pulse width \footnote[1]{enters effective time resolution by convolution}  &  $\sim$~75~fs \\
             X-ray pulse width  \footnotemark[1]{}  &  $\sim$~30~fs \\                  
             remaining jitter after BAM correction \footnotemark[1]{}$^{,}$\footnote[2]{only train-to-train jitter is corrected}  &  $\leq$ 10~fs \\
             \hspace{10pt}   - train-to-train arrival time jitter/drift   &  55~fs \\
             \hspace{10pt} - intra-train shot-to-shot arrival time jitter/drift  &  30~fs \\
             remaining drifts after LAM correction \footnotemark[1]{}  &   $\leq$ 35~fs\\
              X-ray pulse elongation by monochromator  \footnote[3]{enters effective time resolution by summation}  &  $\leq$ 200~fs \\
             \hline
            effective experimental time resolution  & $\leq$ 270~fs 
        \end{tabular}
        \end{ruledtabular}
        \end{table}

Figure \ref{fig:dynamic_XMCD_Fe} displays the transient XMCD  of Fe in the FeNi sample. 
This alloy typically shows a demagnetization time constant of $\sim$~250~fs for Fe in FeNi  depending on the excitation strength and the sample design \cite{janaApplPhysLett2022, Mathias2012PNAS,vonkorffschmisingDirectIndirectExcitation2024, raduSpin2015}. 
Since the time constant is known, the measured transient XMCD allows to experimentally verify the temporal resolution under the present conditions.
We fitted the transient XMCD with a double exponential decay function (one time constant for the demagnetization, the other for the beginning recovery) convoluted by a Gaussian time-resolution function.
Except the demagnetization time which was kept fixed at 250~fs, we left all other parameters free in the fitting process. This yields a FWHM width of the effective time-resolution of 150~$\pm$~30~fs, which overlaps with the lower limit of the temporal resolution estimated above.

\section{\label{conclusions} conclusions } 

We have successfully used ultrashort circularly polarized soft X-ray
pulses at the FLASH2 facility in the photon energy range 700~eV to 860~eV. 
The degree of circular polarization is found to be consistent with 100~\%. 
This photon energy range covers the $L_{3,2}$ edges of the $3d$ elemental magnets Fe and Co, and the $L_{3}$ edge of Ni, which are highly relevant for the field of ultrafast magnetization dynamics.
With the envisaged improvements to the pump laser system to make pulse energies in the train more equal, a higher repetition rate pump laser and improved overall temporal resolution, FLASH will offer even more interesting research possibilities for ultrafast spin dynamics experiments on $3d$ transition metals.

\section{\label{ Acknowledgments } ACKNOWLEDGMENTS }
We acknowledge DESY (Hamburg, Germany), a member of the Helmholtz Association HGF, for the provision of experimental facilities. Parts of this research were carried out at the beamline FL23 at FLASH2. Beamtime was allocated for proposals F-20211728 EC and F-20220710. 
The statics reference measurements were carried out at the PM3 scattering instrument at the BESSY II electron storage ring operated by the Helmholtz-Zentrum Berlin für Materialien und Energie; we would like to thank Torsten Kachel for experimental support.
This work was supported by the Deutsche Forschungsgemeinschaft (DFG) through TRR~227 "Ultrafast Spin Dynamics", ProjectID 328545488 (projects A01, A02 and A03) and by the Bundesministerium für Bildung und Forschung (BMBF) through the project "Spinflash" (05K22KE2).

\section{\label{ Authorcontributions } Author Contributions}

\textbf{S. Marotzke:} Formal analysis (equal); Investigation (equal); Methodology (equal); Software (equal); Visualization (equal); Writing – original draft (equal); Writing – review \&  editing (equal).
\textbf{D. Gupta: } Formal analysis (equal); Investigation (equal); Visualization (equal); Writing – original draft (equal); Writing – review \& editing (equal).
\textbf{R.-P. Wang: } Formal analysis (equal); Investigation (equal); Methodology (equal); Software (equal); Writing – review \& editing (equal).
\textbf{M. Pavelka: } Investigation (equal); Writing – review \& editing (equal).
\textbf{S. Dziarzhytski: } Investigation (equal); Methodology (equal); Writing – review \& editing (equal).
\textbf{C. von Korff Schmising: } Conceptualization (equal); Investigation (equal); Resources (equal); Writing – review \& editing (equal).
\textbf{S. Jana: } Formal analysis (equal); Investigation (equal); Writing – review \& editing (equal).
\textbf{N. Thielemann-Kühn: } Formal analysis (equal); Investigation (equal); Writing – review \& editing (equal).
\textbf{T. Amrhein: } Formal analysis (equal); Investigation (equal); Writing – review \& editing (equal).
\textbf{M. Weinelt: } Funding acquisition (equal); Investigation (equal); Writing – review \& editing (equal).
\textbf{I. Vaskivskyi: } Formal analysis (equal); Investigation (equal); Writing – review \& editing (equal).
\textbf{R. Knut: } Investigation (equal); Writing – review \& editing (equal).
\textbf{D. Engel: } Resources (lead); Writing – review \& editing (equal).
\textbf{M. Braune: } Methodology (equal); Writing – original draft (equal); Writing – review \& editing (equal).
\textbf{M. Ilchen: } Methodology (equal); Writing – review \& editing (equal).
\textbf{S. Savio: } Methodology (equal); Writing – review \& editing (equal).
\textbf{T. Otto: } Methodology (equal); Writing – review \& editing (equal).
\textbf{K. Tiedtke: } Methodology (equal); Writing – review \& editing (equal).
\textbf{V. Scheppe:} Methodology (equal); Writing – review \& editing (equal).
\textbf{J. Rönsch-Schulenberg: } Methodology (equal); Writing – review \& editing (equal).
\textbf{E. Schneidmiller: } Investigation (equal); Methodology (equal); Writing – original draft (equal); Writing – review \& editing (equal).
\textbf{C.~Sch\"u\ss ler-Langeheine: } Conceptualization (equal); Funding acquisition (equal); Investigation (equal); Methodology (equal); Supervision (equal); Writing – review \& editing (equal).
\textbf{H. A. Dürr: } Investigation (equal); Writing – review \& editing (equal).
\textbf{M. Beye: } Conceptualization (equal); Formal analysis (equal); Investigation (equal). Methodology (equal); Project administration (equal); Software (equal); Writing – review \& editing (equal).
\textbf{G. Brenner: } Conceptualization (equal); Investigation (equal); Methodology (equal); Project administration (equal); Visualization (equal); Writing – original draft (equal); Writing – review \& editing (equal).
\textbf{N. Pontius: } Conceptualization (lead); Funding acquisition (equal); Investigation (equal); Methodology (equal); Project administration (equal); Supervision (equal); Visualization (equal); Writing – original draft (lead); Writing – review \& editing (equal).

\section{\label{DataAvailability} DATA AVAILABILITY}

The data that support the findings of this study will be openly available on https://zenodo.org/. Raw data were generated at DESY and BESSY II (HZB). Derived data supporting the findings of this study are available from the corresponding author upon reasonable request.

\newpage

\bibliography{ref}

\end{document}